\documentclass[aps,prd,twocolumn,amsmath,amssymb,superscriptaddress,nofootinbib,showpacs]{revtex4-1}
\usepackage{aas_macros}
\usepackage{graphicx}
\usepackage{bm}
\usepackage{wasysym}

\usepackage[colorlinks=true]{hyperref}

\newcommand{\bs}{\bar s}
\newcommand{\bsl}{\bar s_\textrm{LLR}}
\newcommand{\bsa}{\bar s_\textrm{AI}}
\newcommand{\bS}{\bar S_{\astrosun}}
\newcommand{\ba}[1]{\left(\bar a^{#1}_\textrm{eff}\right)}
\DeclareTextSymbol{\degre}{T1}{6}
\DeclareTextSymbol{\degre}{OT1}{23}

\begin{document}

\def\t{\tilde}

\title{Testing Lorentz symmetry with planetary orbital dynamics}

\author{A.~Hees}
\email{a.hees@ru.ac.za}
\affiliation{Department of Mathematics, Rhodes University, 6140 Grahamstown, South Africa}

\author{Q.~G.~Bailey}
\affiliation{Department of Physics, Embry-Riddle Aeronautical University, 3700 Willow Creek Road, Prescott, AZ 86301, USA}

\author{C.~Le~Poncin-Lafitte}
\affiliation{SYRTE, Observatoire de Paris, PSL Research University, CNRS, Sorbonne Universit\'es, UPMC Univ. Paris 06, LNE, 61 avenue de l'Observatoire, 75014 Paris, France}

\author{A.~Bourgoin}
\affiliation{SYRTE, Observatoire de Paris, PSL Research University, CNRS, Sorbonne Universit\'es, UPMC Univ. Paris 06, LNE, 61 avenue de l'Observatoire, 75014 Paris, France}

\author{A.~Rivoldini}
\affiliation{Royal Observatory of Belgium, Avenue Circulaire, 3, 1180 Bruxelles, Belgium}

\author{B.~Lamine}
\affiliation{IRAP, Universit\'e de Toulouse, UPS-OMP, CNRS, F-31028 Toulouse, France}

\author{F.~Meynadier}
\affiliation{SYRTE, Observatoire de Paris, PSL Research University, CNRS, Sorbonne Universit\'es, UPMC Univ. Paris 06, LNE, 61 avenue de l'Observatoire, 75014 Paris, France}

\author{C.~Guerlin}
\affiliation{Laboratoire Kastler Brossel, ENS-PSL Research University, CNRS, UPMC-Sorbonne Universit\'es, Coll\`ege de France, 24 rue Lhomond, 75005 Paris}
\affiliation{SYRTE, Observatoire de Paris, PSL Research University, CNRS, Sorbonne Universit\'es, UPMC Univ. Paris 06, LNE, 61 avenue de l'Observatoire, 75014 Paris, France}

\author{P.~Wolf}
\affiliation{SYRTE, Observatoire de Paris, PSL Research University, CNRS, Sorbonne Universit\'es, UPMC Univ. Paris 06, LNE, 61 avenue de l'Observatoire, 75014 Paris, France}

\date{August, 2015}


\begin{abstract}
	Planetary ephemerides are a very powerful tool to constrain deviations from the theory of General Relativity using orbital dynamics. The 	effective field theory framework called the Standard-Model Extension (SME) has been developed in order to systematically parametrize hypothetical violations of  Lorentz symmetry (in the Standard Model and in the gravitational sector). In this communication, we use the latest determinations of the supplementary advances of the perihelia and of the nodes obtained by planetary ephemerides analysis to constrain SME coefficients from the pure gravity sector and also from gravity-matter couplings. Our results do not show any deviation from GR and they improve current constraints.  Moreover, combinations with existing constraints from Lunar Laser Ranging and from atom interferometry gravimetry allow us to disentangle contributions from the pure gravity sector from the gravity-matter couplings.
\end{abstract}

\maketitle

\section{Introduction}

The Solar System has proven to be an efficient laboratory to discover new phenomena from gravitational observations. Historically, one can mention the discovery of ``dark'' components (such as the planet Neptune predicted by Le Verrier) or evidence towards non-Newtonian gravity theories (for example the perihelion advance of Mercury which pointed towards General Relativity -- GR). The Solar System remains the most precise laboratory to test the theory of gravity, that is to say GR.

Constraints on deviations from GR can only be obtained in an extended  theoretical framework that parametrizes such deviations. The constraints that are obtained from observations are framework-dependent. In the past decades, two frameworks were widely used in the literature at the scale of the Solar System, namely the Parametrized Post-Newtonian (PPN) formalism~\cite{will:1993fk,*will:2014la} and the fifth force framework~\cite{talmadge:1988uq,*fischbach:1999ly,*adelberger:2009zr}. Stringent constraints have been obtained for these formalisms~\cite{bertotti:2003uq,konopliv:2011dq,lambert:2009bh,*lambert:2011yu,pitjeva:2013fk,verma:2014jk,*fienga:2014uq,williams:2009ys,will:2014la}. More recently, other phenomenological frameworks have been developed like the Standard-Model Extension (SME). The SME is an extensive formalism that allows a systematic description of  Lorentz symmetry violations in all sectors of physics, including gravity~\cite{colladay:1997vn,colladay:1998ys,kostelecky:2004fk}. Violations of Lorentz symmetry are possible in a number of scenarios described in the literature.  While some early motivation came from string theory~\cite{kostelecky:1989yu,*kostelecky:1989jk}, Lorentz violations can also appear in loop quantum gravity, noncommutative field theory and others~\cite{tasson:2014qv,mattingly:2005uq}. The SME is an effective field theory aiming at making phenomenological connections between fundamental theories and experiments.

In particular, a hypothetical Lorentz violation in the gravitational sector naturally leads to an expansion at the level of the action~\cite{kostelecky:2004fk,bailey:2006uq} which in the minimal SME writes
\begin{eqnarray}
	S_\textrm{grav}&=&\int d^4x\frac{\sqrt{-g}}{16\pi G}\left(R-uR+s^{\mu\nu}R^T_{\mu\nu}+t^{\alpha\beta\mu\nu}C_{\alpha\beta\mu\nu}\right) \nonumber\\
	&&\qquad + S'[s^{\mu\nu},t^{\alpha\beta\mu\nu},g_{\mu\nu}]\, ,\label{eq:action}
\end{eqnarray}
with $G$  the gravitational constant, $g$  the determinant of the metric, $R$ the Ricci scalar, $R^T_{\mu\nu}$ the trace-free Ricci tensor, $C_{\alpha\beta\mu\nu}$ the Weyl tensor and $u$, $s^{\mu\nu}$ and $t^{\alpha\beta\mu\nu}$ the Lorentz violating fields.  To avoid conflicts with the underlying Riemann geometry, we assume spontaneous symmetry breaking so that the Lorentz violating coefficients need to be considered as dynamical fields. The last part of the action $S'$ contains the dynamical terms governing the evolution of the SME coefficients. In the linearized gravity limit, the metric depends only on $\bar u$ and $\bar{s}^{\mu\nu}$ which are the vacuum expectation value of $u$ and $s^{\mu\nu}$~\cite{bailey:2006uq}. The coefficient $\bar u$ is unobservable since it can be absorbed in a rescaling of the gravitational constant. The so obtained post-Newtonian metric differs from the one introduced in the PPN formalism~\cite{bailey:2006uq}. In addition to the minimal SME action given by Eq.~(\ref{eq:action}), there exist some higher order Lorentz-violating curvature couplings in the gravity sector (non-minimal SME)~\cite{bailey:2015fk} that have been constrained by short range experiments~\cite{shao:2015uq,*long:2015kx}. These terms are not considered in this communication.

In addition to Lorentz symmetry violations in the pure-gravity sector, violations of Lorentz symmetry can also arise from gravity-matter couplings. In~\cite{kostelecky:2011kx}, it has been shown that gravity-matter couplings violation of Lorentz symmetry can be parametrized by the following classical point mass action
\begin{equation}\label{eq:mat}
	S_\textrm{mat}=\int d\lambda\left(-m\sqrt{g_{\mu\nu}+2c^{\mu\nu}u_\mu u_\nu}-\left( a_\textrm{eff}\right)^\mu u_\mu\right) \, ,
\end{equation}
where $u^\mu$ is the four-velocity of the particle, $m$ is its mass and $c^{\mu\nu}$ and $\left( a_\textrm{eff}\right)^\mu$ are Lorentz violating fields. In this action, spin-coupled Lorentz violation is effectively set to zero. The new fields $c^{\mu\nu}$ and $\left( a_\textrm{eff}\right)^\mu$ depend on the composition of the point particle~\cite{kostelecky:2011kx}. This modification of the action produces two different types of effects: (i) a modification of the way gravity is sourced and (ii) a violation of the three facets of the Einstein Equivalence Principle. The first effect will result in a modification of the space-time metric solution of the field equations. Modifications of the metric in the linearized approximation depend on $\ba{S}$  coefficients, the background values of the coefficients $\left( a_\textrm{eff}\right)^\mu$ from the source body~\cite{kostelecky:2011kx}. On the other hand, the violation of the equivalence principle generated by the action~(\ref{eq:mat}) leads to a deviation from the geodesic motion depending at first order on the coefficients $\bar c_T^{\mu\nu}$ and $\ba{T}^\mu$, the background values of the Lorentz violating fields of the test mass.

Up to now, several studies have constrained the pure-gravity SME coefficients $\bs^{\mu\nu}$ like for example Lunar Laser Ranging~\cite{battat:2007uq}, atom interferometry gravimetry~\cite{muller:2008kx,chung:2009uq}, short range experiment~\cite{bennett:2011vn}, planetary orbital dynamics~\cite{iorio:2012zr}, Gravity Probe B~\cite{bailey:2013kq} and recently binary pulsars~\cite{shao:2014qd,*shao:2014rc}. The  $\ba{w}^\mu$ coefficients are currently poorly constrained by~\cite{hohensee:2011fk,hohensee:2013fp,tasson:2012hl,panjwani:2011qr}. On the opposite, some of the $\bar c^{\mu\nu}$ coefficients are severely constrained (see for example~\cite{wolf:2006sf,hohensee:2011fk,hohensee:2013yg,hohensee:2013fp}). A list of current constraints on all SME coefficients can be found in~\cite{kostelecky:2011ly}. In this study, we will concentrate on the impact of  $\bs^{\mu\nu}$ and $\ba{w}^\mu$ coefficients on planetary orbital dynamics and neglect the $\bar c^{\mu\nu}$ coefficients and leave them for future work. 

In this communication, we show that planetary orbital dynamics can be used to derive stringent constraints on the SME coefficients. Indeed, SME modifications of gravity induce a secular variation of some orbital elements~\cite{bailey:2006uq,kostelecky:2011kx} such as the longitude of the ascending node and the argument of perihelia. These variations are introduced in Sec.~\ref{sec:orbito}. In Sec.~\ref{sec:analysis}, we compare these variations with the present level of residuals coming from INPOP10a (Int\'egrateur Num\'erique Plan\'etaire de l'Observatoire de Paris) ephemerides~\cite{fienga:2011qf}. We use a Bayesian inversion to infer the posterior probability density function (pdf) on the SME coefficients. From the pdf, we estimate correlations between the coefficients. We estimate realistic confidence intervals and also determine linear combinations of the SME coefficients that can be determined independently from planetary orbital dynamics.  In Sec.~\ref{sec:combined}, we combine our results with previous results obtained by Lunar Laser Ranging analysis and atom interferometry gravimetry. Finally, in Sec.~\ref{sec:dis}, we discuss our obtained results and present several ideas that may improve the current analysis.

\section{Effects of SME on orbital dynamics}\label{sec:orbito}
In the linearized gravity limit, the gravity sector of the minimal SME is parametrized by a symmetric trace free tensor $\bs^{\mu\nu}$ and by a scalar $\bar u$ that is unobservable since it corresponds to a rescaling of the gravitational constant~\cite{bailey:2006uq}. Furthermore, the matter-gravity coupling is parametrized amongst others by the $(\bar a_\textrm{eff})^\mu$ coefficients which depend on the composition of the different bodies. The components of these coefficients depend on the observer coordinate system. The standard frame used in the SME formalism labeled by $(T,X,Y,Z)$ is comoving with the Solar System, the spatial axes are defined by equatorial coordinates (see Fig.~1 of~\cite{bailey:2006uq}) and the origin of time is given by the time when the Earth crosses the Sun-centered X-axis at the vernal equinox. The planetary orbital elements are defined with respect to the ecliptic coordinate system. The two coordinate systems differ by a rotation $\mathcal R$ of angle $\varepsilon=23.44$\degre (the Earth obliquity) around the $X$ axis. Therefore, the transformation of the tensor $\bs^{\mu\nu}$ is given by $\bs^{ij}=\mathcal R^i_{\, I}\mathcal R^j_{\, J}\bs^{IJ}$ and $\bs^{0i}=\mathcal R^i_{\, I}\bs^{TI}$ where capital letters refer to the equatorial reference system and lower case letters refer to the ecliptic one. Similarly, the transformation of the $(\bar a_\textrm{eff})^\mu$ vector is given by $\ba{}^{i}=\mathcal R^i_{\, I}\ba{}^{I}$.

SME modifications of gravity induce different types of effects (for an extensive review, see \cite{bailey:2006uq,kostelecky:2011kx}). Two important effects can have implications on planetary ephemerides analysis: effects on the orbital dynamics and effects on the light propagation. Simulations using the Time Transfer Formalism \cite{teyssandier:2008nx,*hees:2014fk,*hees:2014nr} based on the software presented in \cite{hees:2012fk} have shown that only the $\bs^{TT}$ and $\ba{}^T$ coefficients produce a non-negligible effect on the light propagation (while it has impact only at the next post-Newtonian level on the orbital dynamics~\cite{bailey:2006uq,kostelecky:2011kx}). Since in this analysis we concentrate on orbital dynamics, these coefficients are not considered and will be neglected. This can safely be done since the signatures from the $\bs^{TT}$ and $\ba{}^T$ coefficients on the light propagation are similar to the logarithmic standard Shapiro delay, which is not correlated to orbital dynamics effects. 

The equations of motion in the SME formalism are given in \cite{bailey:2006uq,kostelecky:2011kx}. Neglecting the $\bar c_{\mu\nu}$ contributions, the two-body equation of motion reads
\begin{eqnarray}
	\frac{d^2r^j}{dt^2}&=&-\frac{G_NM}{r^3}r^j+\frac{G_NM}{r^3}\left[\bs^{jk}r^k-\frac{3}{2}\bs^{kl}\frac{r^kr^l}{r^2}r^j \right.\nonumber\\
	&&+ 2\frac{\delta m}{M}\left(\bs^{0k}+\sum_{w=e,p,n}\frac{n_2^w}{\delta m}\alpha\ba{w}^k\right)v^kr^j \nonumber \\
	&&\left.- 2\frac{\delta m}{M}\left(\bs^{0j}+\sum_{w=e,p,n}\frac{n_2^w}{\delta m}\alpha\ba{w}^j\right)v^kr^k\right] \label{eq:2body_eom}\, ,
\end{eqnarray}
where $G_N$ is the observed Newton constant, $M=m_1+m_2$ is the total mass of the two bodies, $\delta m=m_2-m_1$ is the difference of the two masses,  $r^j=r_1^j-r_2^j$ is the relative position of the two masses and
\begin{equation}
	n^w_2=N_1^w-N_2^w \, ,
\end{equation}
with $N_{1,2}^w$ the number of particles of species $w$ in the body $1,2$. The coefficient $\bs^{TT}$ is completely unobservable in this context since absorbed in a rescaling of the gravitational constant (see the discussion in~\cite{bailey:2006uq,bailey:2013kq}). The coefficient $\ba{w}^T$ can also be absorbed in a rescaling of the gravitational constant that depends on the composition of each planet~\cite{kostelecky:2011kx}. In this context, one would observe a different $G_N$ with the different planets. Nevertheless, this effect is expected to be very small~\cite{kostelecky:2011kx} and would not produce any supplementary advances of the perihelia and of the nodes and therefore is neglected in this analysis.

In Eq.~(\ref{eq:2body_eom}), the sums on $w$ need to be done on the electrons, protons and neutrons. In the case of a Sun-planet system, we have $M=m_p + m_{\astrosun}\approx m_{\astrosun}$, $\delta m=m_{\astrosun}-m_p\approx m_{\astrosun}$ and $n_2^w=N_p^w-N_{\astrosun}^w\approx -N_{\astrosun}^w$. The fact that we are neglecting $N_p^w$ means that we are neglecting effects produced by the violation of the universality of free fall. Under these assumptions, the equations of motion depend on 
\begin{subequations}\label{eq:bigS}
\begin{eqnarray}
\bS^{0j}&=&	\bs^{0j}-\sum_w \frac{N^w_{\astrosun}}{m_{\astrosun}}\alpha \ba{w}^j  \, ,\\
 &\approx& \bs^{0j}-0.9 \alpha \ba{e+p}^j -0.1 \alpha \ba{n}^j \, ,
\end{eqnarray}
\end{subequations}
 where we used a simple model for the composition of the Sun characterized by $N^e_{\astrosun}/m_{\astrosun}=N^p_{\astrosun}/m_{\astrosun}\approx 0.9\, \textrm{(GeV/c$^2$)}^{-1}$ and $N^n_{\astrosun}/m_{\astrosun}\approx 0.1\, \textrm{(GeV/c$^2$)}^{-1}$ as described in~\cite{kostelecky:2011kx} (with $c$ the speed of light in vacuum). In this paper, $\alpha \ba{w}^j$ is always expressed in GeV/c$^2$ and
\begin{equation}\label{eq:baep}
	\ba{e+p}^j=\ba{e}^j+\ba{p}^j \, .
\end{equation}

Using the Gauss equations, secular perturbations induced by SME on the orbital elements can be computed similarly to what is done in  \cite{bailey:2006uq,iorio:2012zr}. The two orbital elements needed for our analysis are the longitude of the ascending node $\Omega$ and the argument of the perihelion $\omega$. The secular change in these two elements is given by
\begin{subequations}\label{eq:SMEad}
	\begin{eqnarray}
		\left<\frac{d\Omega}{dt}\right>&=&\frac{n}{\sin i(1-e^2)^{1/2}}\left[\frac{\varepsilon}{e^2}\bs_{kP}\sin\omega    \right.\nonumber\\
		&&\left. +\frac{(e^2-\varepsilon)}{e^2}\bs_{kQ}\cos\omega- \frac{2na\varepsilon}{ec}\bS^k \cos\omega\right]\, ,\\
	\left<\frac{d\omega}{dt}\right>&=&	-\cos i \left<\frac{d\Omega}{dt}\right> -n\left[\frac{(e^2-2\varepsilon)}{2e^4}(\bs_{PP}-\bs_{QQ})\right. \nonumber\\
 &&\qquad \left.+\frac{2na(e^2-\varepsilon)}{ce^3(1-e^2)^{1/2}}\bS^Q	\right] \, ,
	\end{eqnarray}
\end{subequations}
where $a$ is the semimajor axis, $e$ the eccentricity, $i$ the orbit inclination (with respect to the ecliptic), $n=(G_Nm_\odot/a^3)^{1/2}$ is the mean motion and $\varepsilon=1-(1-e^2)^{1/2}$. In all these expressions, the coefficients for Lorentz violation with subscripts $P$, $Q$, and $k$ are understood to be appropriate projections of $\bs^{\mu\nu}$ along the unit vectors $P$, $Q$, and $k$, respectively. For example, $\bS^k=k^i \bS^{Ti}$, $\bs_{PP}=P^iP^j \bs^{ij}$. The unit vectors $P$, $Q$ and $k$ define the orbital plane
\begin{subequations}
	\begin{eqnarray}
		\vec P&=&\left(\cos\Omega\cos\omega-\cos i\sin\Omega\sin\omega\right)\vec e_x  \\
		&&+\left(\sin\Omega\cos\omega+\cos i\cos\Omega\sin\omega\right)\vec e_y+\sin i\sin\omega \vec e_z\, ,\nonumber\\
		\vec Q&=&-\left(\cos\Omega\sin\omega + \cos i\sin\Omega\cos\omega\right)\vec e_x  \\
		&&+\left(\cos i\cos\Omega\cos\omega - \sin\Omega\sin\omega\right)\vec e_y+\sin i\cos\omega \vec e_z\, ,\nonumber\\
		\vec k &=&\sin i\sin\Omega	\vec e_x -\sin i\cos\Omega\vec e_y+\cos i \vec e_z \, ,
	\end{eqnarray}
\end{subequations}
where $\vec e_{x,y,z}$ define the basis of the ecliptic reference system. The relations (\ref{eq:SMEad}) are generalizations of Eqs.~(168-171) from \cite{bailey:2006uq} that do not include the $\ba{w}^j$ terms. 

\section{Analysis and results}\label{sec:analysis}
Planetary ephemerides analysis uses an impressive number of different observations to produce high accurate planetary and asteroid trajectories. The observations used to produce ephemerides comprise radioscience observations of spacecraft that orbited around Mercury, Venus, Mars and Saturn, flyby tracking of spacecraft close to Mercury, Jupiter, Uranus and Neptune and optical observations of all planets \cite{folkner:2009fk,konopliv:2011dq,folkner:2010kx,folkner:2014uq,fienga:2008fk,fienga:2009kx,fienga:2011qf,verma:2014jk,fienga:2014uq,pitjeva:2005kx,pitjeva:2013fk,pitjev:2013qv,pitjeva:2014fj}. Estimations of supplementary advances of perihelia with the Russian Ephemerides of Planets and the Moon (EPM) are presented in \cite{pitjeva:2013fk,pitjev:2013qv}. The INPOP ephemerides have produced estimations of supplementary advances of perihelia and nodes. Tab.~\ref{tab:inpop} gives estimations obtained by INPOP10a \cite{fienga:2011qf} on supplementary longitude of nodes $\dot\Omega$ and on supplementary argument of perihelia\footnote{In \cite{fienga:2011qf}, $\dot\omega$ is noted $\dot\varpi$ which is commonly used for the longitude of the perihelion but the estimated values correspond to supplementary argument of perihelia and not to longitude of perihelia (usually noted by $\varpi$) \cite{fienga:2015pr}.} $\dot\omega$.

\begin{table}[htb]
\caption{Values of supplementary longitude of nodes and argument of perihelia estimated by INPOP10a (see Tab.~5 from \cite{fienga:2011qf}). These values are estimated in \cite{fienga:2011qf} as the interval in which the differences of postfit residuals are below 5 \%.}
\label{tab:inpop} % is used to refer this table in the text
\centering
\begin{tabular}{c c c }
\hline
 Planet ~~~& $~~~\dot\Omega$ (mas $\times$ cy$^{-1}$)~~~ & ~~~$\dot\omega$ (mas $\times$ cy$^{-1}$) \\\hline\hline
Mercury & $1.4\pm 1.8$ & $0.4 \pm 0.6$ \\
Venus & $0.2\pm 1.5$ & $0.2 \pm 1.5$ \\
EMB & $0.0\pm 0.9$ & $-0.2 \pm 0.9$ \\
Mars & $-0.05\pm 0.13$ & $-0.04 \pm 0.15$ \\
Jupiter & $-40\pm 42$ & $-41 \pm 42$ \\
Saturn & $-0.1\pm 0.4$ & $0.15 \pm 0.65$ \\
\hline
\end{tabular}
\end{table}

Since $\bar s^{TT}$ and $\ba{w}^T$ do not play any role in the orbital dynamics and $\bar s^{\mu\nu}$ is trace free, the observations depend on 8 independent fundamental coefficients: $\bs^{XX}-\bs^{YY}$, $\bs^Q=\bs^{XX}+\bs^{YY}-2\bs^{ZZ}$, $\bs^{XY}$, $\bs^{XZ}$, $\bs^{YZ}$ and $\bS^{TJ}$ (these coefficients will be denoted as $p_i$ in the following). In this communication, we perform a Bayesian inversion to infer knowledge on these 8 independent coefficients using a Monte Carlo Markov Chain (MCMC) algorithm. The approach is very similar to the one used for binary pulsar data~\cite{shao:2014qd,shao:2014rc}. The observations are assumed to be independent and the errors to be normally distributed. The pdf describing the likelihood (i.e. the probability to obtain observations $O_i$ given certain values of the SME coefficients $p_k$) is given by %$	L(O_i|p_1, p_2, \dots p_n)=\textrm{cst} \, e^{-\chi^2/2}$
\begin{equation}
	L(O_i|p_1, p_2, \dots p_n)=\textrm{cst} \, e^{-\chi^2/2}
\end{equation}
 where the $\chi^2$ is computed by
\begin{eqnarray}
	\chi^2&=&\sum_{pl} \frac{\left(\dot\omega_{pl,\textrm{SME}}(p_k)-\dot\omega_{pl,\textrm{INPOP}}\right)^2}{\sigma^2_{\dot\omega_{pl}}}\\
	&&\qquad+\frac{\left(\dot\Omega_{pl,\textrm{SME}}(p_k)-\dot\Omega_{pl,\textrm{INPOP}}\right)^2}{\sigma^2_{\dot\Omega_{pl}}}\, ,\nonumber
\end{eqnarray}
where the index $pl$ of the sum is running over the six different planets from Tab.~\ref{tab:inpop}, $\dot\Omega_{pl,\textrm{INPOP}}$, $\dot\omega_{pl,\textrm{INPOP}}$ and the corresponding $\sigma$ are from Tab.~\ref{tab:inpop} and where $\dot\omega_{pl,\textrm{SME}}(p_k)$ and $\dot\Omega_{pl,\textrm{SME}}(p_k)$ are simulated values depending on the SME coefficients by (\ref{eq:SMEad}). The posterior pdf of the SME coefficients is given by
\begin{equation}
	P(p_1, p_2, \dots p_n|O_i)=\mathcal C \, L(O_i|p_1,  \dots p_n) \pi(p_1,  \dots p_n) \, ,
\end{equation}
where $\pi(p_1,\dots p_n)=\pi(p_1)\dots \pi(p_n)$ is the prior pdf on the SME coefficients $p_k$ and $\mathcal C$ a constant. We use a uniform prior pdf on the SME coefficients and the MCMC algorithm used is a standard Metropolis-Hasting algorithm~\cite{gregory:2010qv}. We run the Metropolis-Hastings sampler until $10^6$ samples have been generated. The convergence of the MC is ascertained by monitoring the estimated Bayesian confidence intervals of the parameters. Finally, to diminish the effect of the starting configuration, we discard the first 1000 samples.

The marginal pdf of a single SME coefficient $p_j$ is given by 
\begin{equation}
	P(p_j|O_i)= \int dp_1 \int dp_2 \dots P(p_1,\dots, p_n | O_i) \, ,
\end{equation}
where the integrals are performed over all the SME coefficients $p_k$ except $p_j$.

\begin{figure}[htb]
\centering
\includegraphics[width=.98\linewidth]{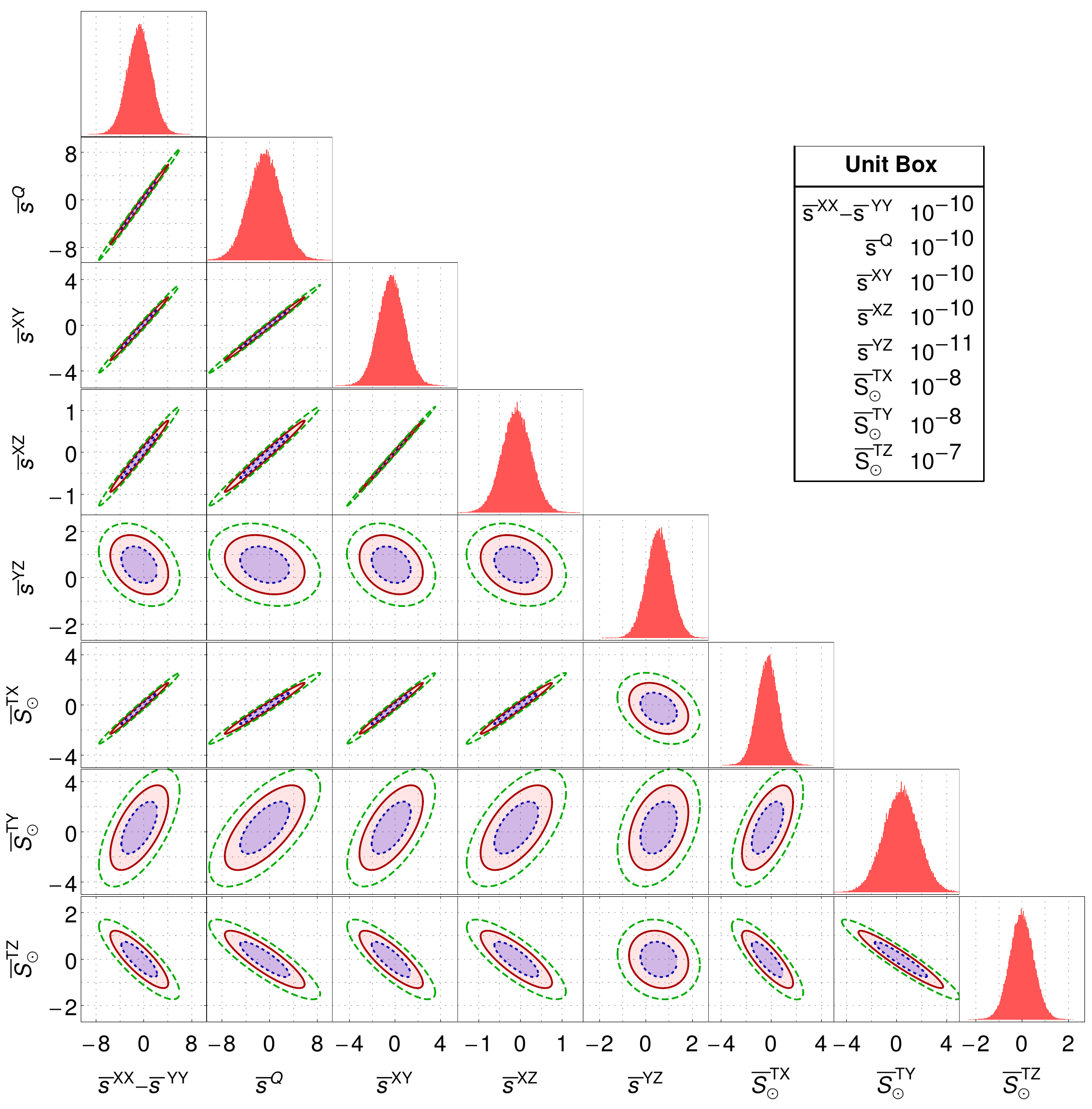}
\caption{2D marginal posterior pdf (useful to assess the correlations).  On the 2D plots, the blue dotted contours represent the 67 \% Bayesian confidence area, the red continuous contour represent the 95 \% Bayesian confidence area and the dashed green contours represent the 99.7 \% Bayesian confidence area. The histograms represent the marginal pdf of the SME coefficients.}
\label{fig:hist1}
\end{figure}

A first run shows that the coefficients of our model are highly correlated, see Fig.~\ref{fig:hist1}. We have used the correlation matrix estimator to assess the strength of the parameters correlations, see Tab.~\ref{tab:corr}. These correlations are mainly due to the fact that all planets have very similar, low inclination, orbital planes. Nevertheless, we can produce marginal 1D posterior distribution for each of the 8 SME coefficients. The histograms corresponding to these distributions are presented in Fig.~\ref{fig:hist1}. The corresponding Bayesian confidence intervals are presented in Tab.~\ref{tab:results}.

\begin{table}[htb]
\caption{Estimations of the SME coefficients. These estimations are still correlated and the correlation matrix is given in Tab.~\ref{tab:corr}. The uncertainties correspond to the 68\% Bayesian confidence levels of the marginal pdf.}
\label{tab:results} % is used to refer this table in the text
\centering
\begin{tabular}{c r  }
\hline
 SME coefficients &  \multicolumn{1}{c}{Estimation} \\\hline\hline
 $\bs^{XX}-\bs^{YY}$			 & $(-0.8 \pm 2.0)\times 10^{-10}$ \\
 $\bs^Q=\bs^{XX}+\bs^{YY}-2~\bs^{ZZ} $ 	 & $(-0.8 \pm 2.7)\times 10^{-10}$\\
 $\bs^{XY}$        	 & $(-0.3 \pm 1.1)\times 10^{-10}$\\
 $\bs^{XZ}$ 		 & $(-1.0 \pm 3.5)\times 10^{-11}$\\
 $\bs^{YZ}$ 		 & $(5.5 \pm 5.2)\times 10^{-12}$\\
 $\bS^{TX}$	 	& $(-2.9 \pm 8.3)\times 10^{-9\phantom{1}}$ \\
 $\bS^{TY}$		 & $(0.3 \pm 1.4)\times 10^{-8\phantom{1}}$ \\
 $\bS^{TZ}$ 		 & $(-0.2 \pm 5.0)\times 10^{-8\phantom{1}}$ \\
\hline
\end{tabular}
\end{table}

\begin{table}[htb]
\caption{Estimations of the correlations coefficients between the different SME coefficients: $\bs^{XX}-\bs^{YY}$, $\bs^Q$, $\bs^{XY}$, $\bs^{XZ}$, $\bs^{YZ}$, $\bS^{TX}$,  $\bS^{TY}$ and $\bS^{TZ}$.}
\label{tab:corr} % is used to refer this table in the text
\centering
\begin{tabular}{c c c c c c c c}\hline\hline
1\\
 \phantom{-}0.99 & 1\\
 \phantom{-}0.99 & \phantom{-}0.99   & 1 \\
 \phantom{-}0.98   & \phantom{-}0.98 & \phantom{-}0.99 &1\\
	-0.32          & -0.24           & -0.26           & -0.26            &1\\
 \phantom{-}0.99  & \phantom{-}0.98  & \phantom{-}0.98 & \phantom{-}0.98  & -0.32            &1\\
 \phantom{-}0.62  & \phantom{-}0.67  & \phantom{-}0.62  & \phantom{-}0.59 & \phantom{-}0.36  & \phantom{-}0.60&1\\
  -0.83          & -0.86             & -0.83           & -0.81            & -0.14             &-0.82 & -0.95 &1\\
\hline\hline
\end{tabular}
\end{table}

\begin{figure}[htb]
\centering
\includegraphics[width=.98\linewidth]{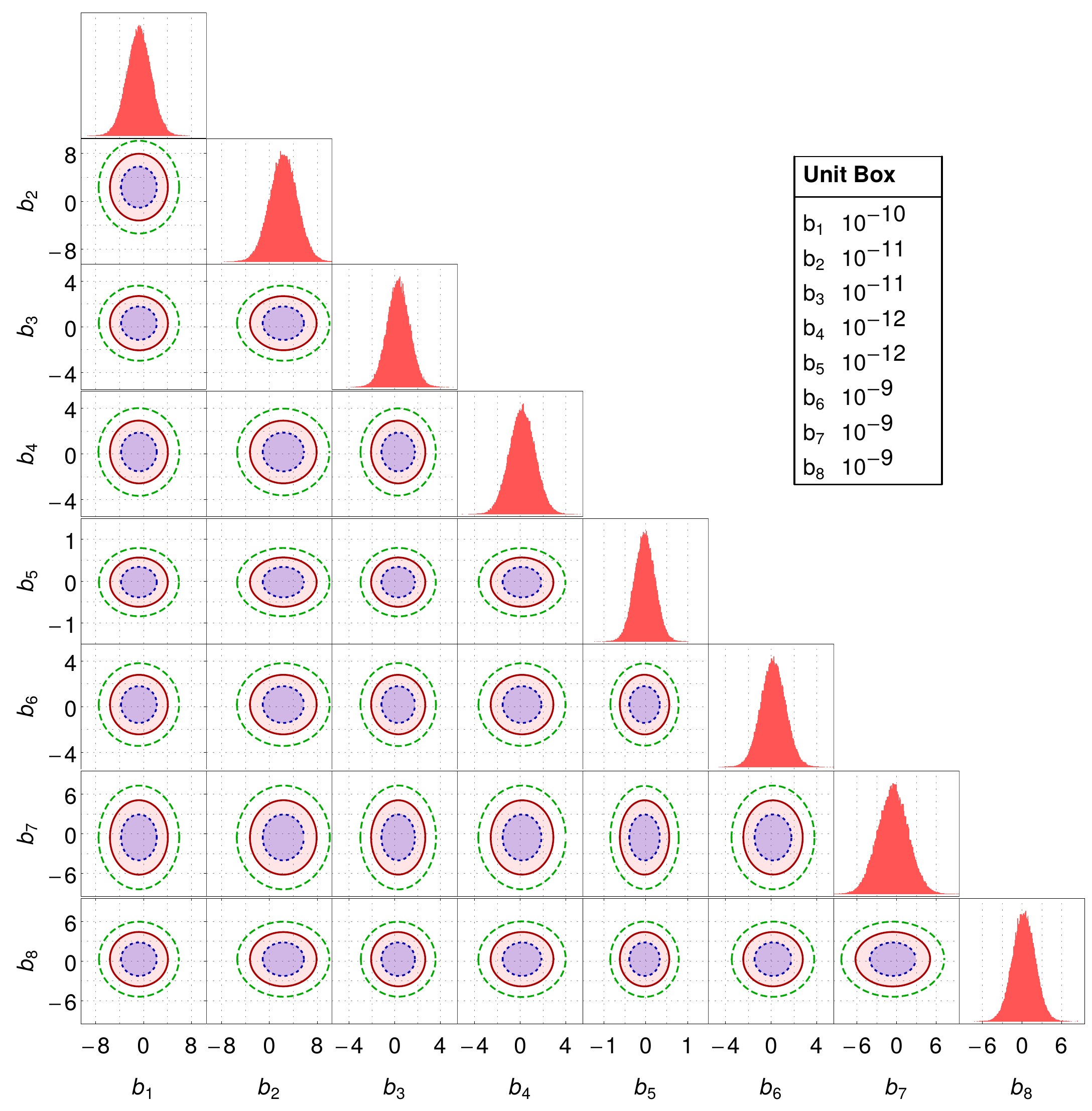}
\caption{2D marginal posterior pdf (useful to assess the correlations) of the linear combinations $b_i$ of the SME coefficients given by Eqs.~(\ref{eq:lin_ind}). On the 2D plots, the blue dotted contours represent the 67 \% Bayesian confidence area, the red continuous contour represent the 95 \% Bayesian confidence area and the dashed green contours represent the 99.7 \% Bayesian confidence area. The 1D histograms represent the  marginal pdf of the SME linear combinations $b_i$.}
\label{fig:hist2}
\end{figure}

Another approach (based on the first run) to avoid highly correlated coefficients is to find the independent linear combinations of the SME coefficients that can be determined by planetary ephemerides analysis. This can be done numerically by performing a normalized Cholesky decomposition of the covariance matrix
\begin{equation}
	C=K^T D^2 K \, ,
\end{equation}
where $C$ is the covariance matrix of the SME coefficients estimated from our first run, $K$ is an upper triangular matrix whose diagonal elements are unity and $D$ is a diagonal matrix. Then the linear combinations $\bm b$ of the \emph{fundamental} SME coefficients (noted $\bm p$) given by
\begin{equation}
	\bm b = K^{-T}\bm p \, ,
\end{equation} 
with $K^{-T}$ the inverse of the transpose of $K$, can be determined completely independently by the analysis of planetary orbital dynamics. In our case, this Cholesky decomposition ($K^{-T}$) is given by
\begin{subequations}\label{eq:lin_ind}
	\begin{eqnarray}
		b_1&=&\left(\bs^{XX}-\bs^{YY}\right) \, ,\\
		b_2&=&-1.37 b_1 + \bs^Q \, ,\\
		b_3&=& -0.15 b_1 -0.31 \bs^Q + \bs^{XY} \, , \\
		b_4&=& 0.013 b_1 +0.064 \bs^Q  -0.48 \bs^{XY} + \bs^{XZ} \, , \\
		b_5&=& 0.26 b_1 -0.31 \bs^Q +0.81 \bs^{XY} -1.67 \bs^{XZ}\nonumber\\
		&& \quad + \bs^{YZ} \, \\
		b_6 &=& -35.5 b_1 + 9.35 \bs^Q - 22.67 \bs^{XY} - 33.95 \bs^{XZ} \nonumber \\
		&& \quad + 7.83 \bs^{YZ} + \bS^{X} \,  ,\\
		b_7&=& 1641.4 b_1 - 2101.1 \bs^Q +4939.9 \bs^{XY} -8846.8\bs^{XZ} \nonumber \\
		&&\quad + 4810.6 \bs^{XZ} -0.89 \bS^X + \bS^Y \, , \\
		b_8&=& 44.5 b_1 + 47.1 \bs^Q -580.1  \bs^{XY} +1041.3 \bs^{XZ} \nonumber\\
		 &&\quad+ 231.5 \bs^{YZ} +3.43  \bS^X +2.56\bS^Y +\bS^Z \, ,
	\end{eqnarray}
\end{subequations}
with the expression of $\bS^J$ given by Eq.~(\ref{eq:bigS}). We can now use the  linear combinations $b_i$ as fundamental parameters for our analysis. Performing a new MC run (using the same prior and likelihood as previously), we show that they can be estimated without any correlation. This can be seen in Fig.~\ref{fig:hist2} where the 2D marginal posterior pdf on the $b_i$ combinations are presented. More quantitatively, the computation of the correlation matrix shows that the $b_i$ combinations are completely decorrelated by planetary ephemerides analysis since the absolute values of the correlation parameters never exceed 0.03. The 1D posterior pdf of the $b_i$ combinations are also represented in Fig.~\ref{fig:hist2}. The estimated mean and standard deviation are given in Tab.~\ref{tab:results2}. The obtained uncertainties are much smaller than those given in Tab.~\ref{tab:results}. 

\begin{table}[htb]
\caption{Estimations of the independent linear combinations $b_i$ of the SME coefficients. The expressions of the combinations $b_i$ are given by Eqs.~(\ref{eq:lin_ind}). The uncertainties correspond to the 68\% Bayesian confidence levels of the marginal pdf.}
\label{tab:results2} % is used to refer this table in the text
\centering
\begin{tabular}{c r  }
\hline
 SME linear combinations &  \multicolumn{1}{c}{Estimation} \\\hline\hline
 $b_1$	 & $(-0.8 \pm 2.0)\times 10^{-10}$ \\
 $b_2$ 	 & $(2.3 \pm 2.3)\times 10^{-11}$\\
 $b_3$   & $(3.0 \pm 9.7)\times 10^{-12}$\\
 $b_4$ 	 & $(0.2 \pm 1.1)\times 10^{-12}$\\
 $b_5$ 	 & $(-0.3 \pm 2.4)\times 10^{-13}$\\
 $b_6$	 & $(0.2 \pm 1.1)\times 10^{-9\phantom{1}}$ \\
 $b_7$	 & $(-0.6 \pm 2.3)\times 10^{-9\phantom{1}}$ \\
 $b_8$ 	 & $(0.3 \pm 1.7)\times 10^{-9\phantom{1}}$ \\
\hline
\end{tabular}
\end{table}

We want to emphasize the fact that the results from both approaches presented above are completely equivalent. They are two ways to represent the same results. One is free to choose which approach is more appropriate: to work with the fundamental SME coefficients determined by Tab.~\ref{tab:results} at the price of including the covariance matrix (or equivalently the correlation matrix from Tab.~\ref{tab:corr}) in the analysis or to work with uncorrelated linear combinations of the SME coefficients that are determined by Tab.~\ref{tab:results2}. The results provided by both approaches describe the same physical information. Therefore, they are completely equivalent.

\section{Combination with Lunar Laser Ranging and atom interferometry gravimetry}\label{sec:combined}
It is interesting to combine the results obtained in the last section with constraints available in the literature. In particular, Lunar Laser Ranging (LLR) data have been used to constrain the pure gravity sector of SME~\cite{battat:2007uq}. Similarly, atomic gravimetry data have also been used to constrain the $\bs^{\mu\nu}$ coefficients~\cite{muller:2008kx,chung:2009uq}. We will first combine our results from Sec.~\ref{sec:analysis} with LLR results to produce constraints on the SME pure gravity sector alone. This will highlight the improvement brought by the planetary ephemerides data. In a second step, we will consider both the pure gravity sector and the gravity-matter couplings coefficients. We will demonstrate that the combination of planetary ephemerides data, LLR data and atom interferometry gravimetry data allows one to completely disentangle all the SME coefficients $\bs^{\mu\nu}$ and $\ba{w}^J$.

The procedure to combine different types of analysis is standard and consists of performing a global least squares fit of all the estimations available. Obviously, the planetary estimations given by Tab.~\ref{tab:results} are not independent. To take into account the correlation between the coefficients estimated in Sec.~\ref{sec:analysis}, we use the parameter covariance matrix from Tab.~\ref{tab:corr} as a weight in the least squares fit. Similarly, the coefficients estimated in the LLR analysis are weighted by their standard deviation in the least squares fit. Since no covariance matrix can be found in the literature, we assume these estimations to be independent (this corresponds to a worst case scenario). Instead of working with results given in Tab.~\ref{tab:results} that are correlated, we can equivalently  use the linear combinations given by Eqs.~(\ref{eq:lin_ind}) and we then use the estimated standard deviations from Tab.~\ref{tab:results2} to weight the least squares fit. In that approach, the weight matrix in the fit is diagonal. We insist on the fact that both approaches lead to the same results. In the following we provide the mean and the standard deviation of the SME coefficients as given from the least square fit.

\subsection{Pure gravity sector}
First, let us focus on the pure gravity sector alone and neglect the $\ba{w}^J$ coefficients. It has been shown in \cite{bailey:2006uq} that the main oscillations in the radial distance between the Earth and the Moon due to the $\bs^{\mu\nu}$ coefficients depend on 6 linear combinations: $\bs^{11}-\bs^{22}$, $\bs^{12}$, $\bs^{01}$, $\bs^{02}$, $\bs_{\Omega_\oplus c}$ and $\bs_{\Omega_\oplus s}$. They can be expressed in terms of the standard  SME coefficients expressed in an Earth equatorial frame and in terms of the longitude of the ascending node $\alpha$ and of the inclination $\beta$ of the Moon's orbit with respect to the equator. These combinations are given by Eqs.~(107-108) from \cite{bailey:2006uq}. The longitude of the ascending node $\alpha$ with respect to the equator oscillates around 0. This oscillation is due to the secular advance of the longitude of the ascending node with respect to the ecliptic. Similarly, the inclination of the Moon's orbit with respect to the equator oscillates around $\beta=23.44$\degre. As a consequence, the transformation of the LLR linear combinations to the standard SME coefficients is given by
\begin{subequations}\label{eq:lin_LLR}
	\begin{eqnarray}
			\bsl^A&=&\bs^{11}-\bs^{22}=0.92\left(\bs^{XX}-\bs^{YY}\right) \label{eq:bsla}\\
			&& +0.08 \left(\bs^{XX}+\bs^{YY}-2\bs^{ZZ}\right)-0.73\bs^{YZ}\, ,\nonumber\\
			\bsl^B&=&\bs^{12}=0.92\bs^{XY}+0.40\bs^{XZ} \, ,\label{eq:bslb}\\
			\bsl^C&=&\bs^{02}= 0.92\bs^{TY} +0.40\bs^{TZ}\, , \label{eq:sLLRc}\\
			\bsl^D&=&\bs^{01}=\bs^{TX}   \, , \\
			\bsl^E&=&\bs_{\Omega_\oplus c}=-3.21\bs^{TY}-1.39\bs^{TZ}\, , \label{eq:somega_earth_c}\\
			\bsl^F&=&\bs_{\Omega_\oplus s}=-3.50\bs^{TX} \, .\label{eq:somega_earth_s}
	\end{eqnarray}
\end{subequations}
Note that the above transformations are different from those used in \cite{chung:2009uq}. In that paper, the
authors have used $\alpha=125$\degre, which corresponds to the transformation between the lunar plane and the ecliptic plane at the date J2000 while the reference frame used in the SME framework is the equatorial plane (and not the ecliptic one). Therefore, the value of $\alpha$ and $\beta$ needs to be taken with respect to the equatorial plane at the moment where the experiment was performed, or as their average value if they vary during the experiment.\footnote{Note that \cite{bailey:2006uq} advised caution on this point: ``For definiteness and to acquire insight, we adopt the values $\alpha=125$\degre  and $\beta=23.5$\degre.  However, these angles vary for the Moon due to comparatively large Newtonian perturbations, so some caution is needed in using the equations that follow.''}

In \cite{battat:2007uq}, Battat \emph{et al} have fitted the amplitudes related to the signature of the 6 SME combinations (\ref{eq:lin_LLR}) on residuals of LLR analysis. As a result, they obtained constraints given in Tab.~\ref{tab:LLR}.
\begin{table}[htb]
\caption{Estimations of the SME coefficients derived from LLR analysis from \cite{battat:2007uq}.}
\label{tab:LLR} % is used to refer this table in the text
\centering
\begin{tabular}{c r  }
\hline
 SME linear combination &  \multicolumn{1}{c}{Estimation} \\\hline\hline
 $\bsl^A$			 & $(1.3 \pm 0.9)\times 10^{-10}$ \\
 $\bsl^B$ 	 & $(6.9 \pm 4.5)\times 10^{-11}$\\
 $\bsl^C$ 		 & $(-5.2 \pm 4.8)\times 10^{-7\phantom{1}}$\\
 $\bsl^D$						 & $(-0.8 \pm 1.1)\times 10^{-6\phantom{1}}$ \\
 $\bsl^E$						 & $(0.2 \pm 3.9)\times 10^{-7\phantom{1}}$ \\
$\bsl^F$ 		 & $(-1.3 \pm 4.1)\times 10^{-7\phantom{1}}$ \\
\hline
\end{tabular}
\end{table}

Combining these constraints with those obtained in the previous section from planetary ephemerides lead to estimations of the pure gravity SME coefficients given in Tab.~\ref{tab:resINPOPLLR}. One can see that the $\bs^{XX}-\bs^{YY}$ and the three coefficients $\bs^{JK}$ (with $J\neq K$) are improved by the combinations of the data. This is mainly due to the fact that the correlations are reduced. It is also worth mentioning that this combined analysis improves the combined  LLR and atom interferometry gravimetry analysis from \cite{chung:2009uq} by 2 to 3 orders of magnitude. 

\begin{table}[htb]
\caption{Estimated mean and 1$\sigma$ uncertainty of the SME coefficients $\bs^{\mu\nu}$ by combining planetary ephemerides analysis from Sec.~\ref{sec:analysis} and LLR analysis~\cite{battat:2007uq}. It has been assumed that the $\ba{w}^J$ coefficients vanish.}
\label{tab:resINPOPLLR} % is used to refer this table in the text
\centering
\begin{tabular}{c r  }
\hline
 SME coefficients &  \multicolumn{1}{c}{Estimation} \\\hline\hline
 $\bs^{XX}-\bs^{YY}$			 & $(9.6 \pm 5.6)\times 10^{-11}$ \\
 $\bs^Q=\bs^{XX}+\bs^{YY}-2~\bs^{ZZ} $ 	 & $(1.6 \pm 0.78)\times 10^{-10}$\\
 $\bs^{XY}$        	 & $(6.5 \pm 3.2)\times 10^{-11}$\\
 $\bs^{XZ}$ 		 & $(2.0 \pm 1.0)\times 10^{-11}$\\
 $\bs^{YZ}$ 		 & $(4.1 \pm 5.0)\times 10^{-12}$\\
 $\bs^{TX}$	 	& $(4.3 \pm 2.5)\times 10^{-9\phantom{1}}$ \\
 $\bs^{TY}$		 & $(1.1 \pm 1.1)\times 10^{-8\phantom{1}}$ \\
 $\bs^{TZ}$ 		 & $(-3.8 \pm 3.0)\times 10^{-8\phantom{1}}$ \\
\hline
\end{tabular}
\end{table}

\subsection{Gravity sector and matter-gravity couplings}
In order to use LLR analysis to constrain simultaneously the $\bs^{\mu\nu}$ and $\ba{w}^J$ coefficients, we need to identify the contributions of the $\ba{w}^J$ coefficients to the amplitudes of the Earth-Moon distance oscillations. The SME contribution to the equations of motion of the Moon-Earth system can be found in \cite{bailey:2006uq,kostelecky:2011kx} and is given by
\begin{widetext}
\begin{eqnarray}
	\left.\frac{d^2 r^J}{dt^2}\right|_\textrm{SME}&=&\frac{G_NM}{r^3} \left[\bs^{JK} r^K - \frac{3 \bs^{KL}r^K r^L r^J}{2r^2} + 3\left(\bar s^{TK}-\sum_{w=e,p,n}\frac{2}{3}\frac{n_3^w}{M} \alpha \ba{w}^K\right)V^K r^J-V^Kr^K\bar s^{TJ} - V^J\bar s^{TK}r^K \right.\label{eq:LLR_eom} \\
&+&\left. \frac{3V^K\bar s^{TL}r^Kr^Lr^J}{r^2} + 2\frac{\delta m}{M}\left( \bar s^{TK}+ \sum_{w=e,p,n} \frac{n_2^w}{\delta m} \alpha \ba{w}^K\right) v^Kr^J -2\frac{\delta m}{M}\left( \bar s^{TJ} + \sum_{w=e,p,n} \frac{n_2^w}{\delta m}\alpha \ba{w}^J \right) v^K r^K  \right] \, ,\nonumber
\end{eqnarray}
\end{widetext}
where $M=m_{\leftmoon}+m_\oplus$, $\delta m=m_\oplus-m_{\leftmoon}\approx m_\oplus$, $r^J$ is the position of the Moon with respect to the Earth, $v^J$ is the relative velocity of the Moon with respect to the Earth, $V^K$ is the heliocentric velocity of the Earth-Moon Barycenter and
\begin{subequations}
	\begin{eqnarray}
		n_2^w&=&N^w_{\leftmoon}-N^w_\oplus \approx-N^w_\oplus  \, ,\\
		n_3^w&=&M\left(\frac{N^w_{\leftmoon}}{m_{\leftmoon}}+\frac{N^w_\oplus}{m_\oplus}\right) \, ,
	\end{eqnarray}
\end{subequations}
where $N^w_i$ is the number of particles of species $w$ in the body $i$. Following the approach described in Appendix A of \cite{bailey:2006uq} (see also~\cite{nordtvedt:1995nr,nordtvedt:1996sf}), we expand the equations of motion around a reference circular orbit and perform a Fourier analysis to obtain the contributions of the $\ba{w}^J$ terms to the oscillations of the Earth-Moon distance. The term proportional to $V^K r^J$ in the first line of Eq.~(\ref{eq:LLR_eom}) leads to an oscillation at the Earth orbital frequency $\Omega_\oplus$. The $\ba{w}^J$ coefficient modifies the expression of $\bs_{\Omega_\oplus,1}$ in Eq.~(A20) from \cite{bailey:2006uq}. Similarly, the modifications of the terms proportional to $\delta m$ in Eq.~(\ref{eq:LLR_eom}) change the expression for $\bs^{01}$ and $\bs^{02}$. To summarize, we find that the $\ba{w}^J$ coefficients will modify the combinations appearing in LLR oscillations as ($\bsl^A$ and $\bsl^B$ being unchanged)
\begin{subequations}\label{eq:llr_approx}
	\begin{eqnarray}
		\bsl^{C}&=&\bs^{02}+\sum_w \frac{n_2^w}{\delta m} \alpha \ba{w}^2\nonumber\\
			&\approx& \bs^{02}- \sum_w \frac{N_\oplus^w}{m_\oplus} \alpha \ba{w}^2 \, , \\
		\bsl^{D}&=&\bs^{01}+\sum_w \frac{n_2^w}{\delta m} \alpha \ba{w}^1\nonumber\\
			&\approx& \bs^{01}- \sum_w \frac{N_\oplus^w}{m_\oplus} \alpha \ba{w}^1 \, , \\
		\bsl^E&=&\bs_{\Omega_\oplus c}+ 2 \sum_w \frac{n_3^w}{M} \left(\cos\eta   \alpha\ba{w}^Y +\sin\eta \alpha\ba{w}^Z \right)\nonumber\\
		&=&\bs_{\Omega_\oplus c}+ 2 \sum_w \left(\frac{N^w_{\leftmoon}}{m_{\leftmoon}}+\frac{N^w_\oplus}{m_\oplus}\right) \left(\cos\eta   \alpha\ba{w}^Y\right. \nonumber\\
		&&\qquad \qquad \qquad \qquad \qquad \quad\left. +\sin\eta \alpha\ba{w}^Z \right) \, , \\
		\bsl^F&=&\bs_{\Omega_\oplus s}+2 \sum_w \frac{n_3^w}{M}   \alpha\ba{w}^X \nonumber \\
		&=&\bs_{\Omega_\oplus s}+2 \sum_w \left(\frac{N^w_{\leftmoon}}{m_{\leftmoon}}+\frac{N^w_\oplus}{m_\oplus}\right)   \alpha\ba{w}^X \, ,
	\end{eqnarray}
\end{subequations}
where $\bs_{\Omega_\oplus c}$ and $\bs_{\Omega_\oplus s}$ are given by Eq.~(108) of \cite{bailey:2006uq} or by Eqs.~(\ref{eq:somega_earth_c}-\ref{eq:somega_earth_s}). A simple model for the composition of the Earth leads to $N^e_{\oplus}/m_{\oplus}=N^p_{\oplus}/m_{\oplus}\approx N^n_{\oplus}/m_{\oplus}\approx 0.5\, \textrm{(GeV/c$^2$)}^{-1}$~\cite{kostelecky:2011kx}. Similarly, the model for the composition of the Moon from~\cite{lodders:1998fk} leads to $N^e_{\leftmoon}/m_{\leftmoon}=N^p_{\leftmoon}/m_{\leftmoon}\approx N^n_{\leftmoon}/m_{\leftmoon}\approx 0.5\, \textrm{(GeV/c$^2$)}^{-1}$. Using these values, the combinations (\ref{eq:sLLRc}-\ref{eq:somega_earth_s}) appearing in LLR data analysis are modified by the $\ba{w}^J$ coefficients as follow:
\begin{subequations}\label{eq:lin_LLR2}
	\begin{eqnarray}
%&&			\bs^{11}-\bs^{22}=0.92\left(\bs^{XX}-\bs^{YY}\right) \\
%			&& \qquad +0.08 \left(\bs^{XX}+\bs^{YY}-2\bs^{ZZ}\right)-0.73\bs^{YZ}\, ,\nonumber\\
			%
%&&			\bs^{12}=0.92\bs^{XY}+0.40\bs^{XZ} \, ,\\
			\bsl^C&=& 0.92\left(\bs^{TY}-0.5\alpha\ba{e+p}^Y-0.5\alpha\ba{n}^Y\right)\\
&&			\quad+0.4\left(\bs^{TZ}-0.5\alpha\ba{e+p}^Z-0.5\alpha\ba{n}^Z\right)\, ,\nonumber \\
			\bsl^D&=&\bs^{TX} -0.5\alpha\ba{e+p}^X-0.5\alpha\ba{n}^X  \, , \\
			\bsl^E&=&-3.21\bs^{TY}-1.39\bs^{TZ} + 1.84\alpha\ba{e+p}^Y \\
&& \qquad + 1.84\alpha\ba{n}^Y +0.8\alpha\ba{e+p}^Z+0.8 \alpha\ba{n}^Z \, , \nonumber \\
			\bsl^F&=&-3.50\bs^{TX} +2\alpha\ba{e+p}^X+2\alpha\ba{n}^X  \, .
	\end{eqnarray}
\end{subequations}

Atom interferometry gravimetry has also been used to constrain SME coefficients \cite{muller:2008kx,chung:2009uq}. A violation of Lorentz symmetry induces periodic variations of the local acceleration that can be measured by atom gravimetry. Amplitudes of these oscillations have been partially computed in \cite{bailey:2006uq} for the $\bar s^{\mu\nu}$ coefficients (see Table IV) and in \cite{kostelecky:2011kx} for the $\ba{w}^J$ coefficients (see Table IV). An improved calculation shows that the $\ba{w}^J$ coefficients modify only two of the amplitudes constrained in \cite{muller:2008kx,chung:2009uq}:
\begin{subequations}
	\begin{eqnarray}
		C_\omega&=&\frac{i_4}{2}\bs^{XZ}\sin 2\chi -2 \frac{V_L}{c} i_5 \bs^{TY} \nonumber \\  
		&& \qquad +\frac{4V_L}{3c}\sum_{w=e,p,n}\left[i_\oplus\frac{N^w_T}{m_T}+\frac{3}{2}\frac{N^w_\oplus}{m_\oplus}\right]\alpha\ba{w}^Y \nonumber \\
		&=&\frac{i_4}{2}\bsa^A \sin 2\chi \, , \\
		D_\omega &=&\frac{i_4}{2}\bs^{YZ}\sin2\chi + 2 \frac{V_L}{c} i_5 \bs^{TX}\nonumber \\
		&&\qquad  -\frac{4V_L}{3c}\sum_{w=e,p,n}\left[i_\oplus\frac{N^w_T}{m_T}+\frac{3}{2}\frac{N^w_\oplus}{m_\oplus}\right]\alpha\ba{w}^X \nonumber \\
		&=&\frac{i_4}{2}\bsa^{B}\sin2\chi \, ,
	\end{eqnarray}
\end{subequations}
where $i_\oplus=I_\oplus/(m_\oplus R_\oplus^2)\approx1/2$  (with $I_\oplus$ the Earth spherical inertial moment and $R_\oplus$ the Earth radius), $i_4=1-3i_\oplus\approx -1/2$, $i_5=1+2i_\oplus/3\approx 4/3$, the subscripts $T$ refer to the test body, $V_L=\omega_\oplus R_\oplus \sin \chi$ is the velocity of the laboratory due to Earth rotation ($\omega_\oplus$ being the angular velocity of the Earth rotation) and $\chi$ is the geographical colatitude of the location where the experiment is performed. In the last expressions, we introduced two linear combinations given by
\begin{subequations}
	\begin{eqnarray}
		\bsa^A&=&\bs^{XZ}-\frac{4}{i_4\sin 2\chi}\frac{V_L}{c}i_5 \bs^{TY}\\
		&&+\frac{8V_L}{3c}\frac{1}{i_4\sin 2 \chi}\sum_{w=e,p,n}\left[i_\oplus\frac{N^w_T}{m_T}+\frac{3}{2}\frac{N^w_\oplus}{m_\oplus}\right]\alpha\ba{w}^Y \nonumber  \, ,\\
%		&=&\bs^{XY}+ICI \, ,\\
		%
		\bsa^B &=& \bs^{YZ}+\frac{4}{i_4\sin 2\chi}\frac{V_L}{c}i_5 \bs^{TX}\\
		&&-\frac{8V_L}{3c}\frac{1}{i_4\sin 2 \chi}\sum_{w=e,p,n}\left[i_\oplus\frac{N^w_T}{m_T}+\frac{3}{2}\frac{N^w_\oplus}{m_\oplus}\right]\alpha\ba{w}^X \nonumber \, .
	\end{eqnarray}
\end{subequations}
For the experiment performed by \cite{muller:2008kx,chung:2009uq}, we have $\chi=42.3$\degre and $V_L/c\approx 1.04 \times 10^{-6}$. Moreover, numerical estimations for a Caesium atom interferometer lead to $N^e_\textrm{Cs}/m_\textrm{Cs}=N^p_\textrm{Cs}/m_\textrm{Cs}=0.44 \, \textrm{(GeV/c$^2$)}^{-1}$, $N^n_\textrm{Cs}/m_\textrm{Cs}=0.63 \, \textrm{(GeV/c$^2$)}^{-1}$. Finally, the values for the Earth are given in \cite{kostelecky:2011kx} and are mentioned above after Eq.~(\ref{eq:llr_approx}). Using these values gives
\begin{subequations}\label{eq:lin_at}
	\begin{eqnarray}
		\bsa^A&=&\bs^{XZ}+1.12\times 10^{-5}\bs^{TY}  \\
		&&\qquad  -5.43 \times 10^{-6}\alpha\ba{e+p}^Y-5.96 \times 10^{-6}\alpha\ba{n}^Y  \nonumber \, ,\\
%		&=&\bs^{XY}+ICI \, ,\\
		%
		\bsa^B &=& \bs^{YZ}-1.12\times 10^{-5}\bs^{TX}  \\
		&&\qquad  +5.43 \times 10^{-6}\alpha\ba{e+p}^X+5.96 \times 10^{-6}\alpha\ba{n}^X  \nonumber \, ,
	\end{eqnarray}
\end{subequations}
with $\ba{e+p}^J$ given by Eq.~(\ref{eq:baep}).

Therefore, the experiment from \cite{muller:2008kx,chung:2009uq} is sensitive to the last two combinations and not to $\bs^{XZ}$ and $\bs^{YZ}$ alone. The results from \cite{chung:2009uq} are presented in Tab.~\ref{tab:atom}

\begin{table}[htb]
\caption{Estimations of the SME coefficients derived from atom interferometry gravimetry by \cite{chung:2009uq,muller:2008kx}.}
\label{tab:atom} % is used to refer this table in the text
\centering
\begin{tabular}{c r  }
\hline
 SME linear combination &  \multicolumn{1}{c}{Estimation} \\\hline\hline
 $\bs^{XX}-\bs^{YY}$			 & $(4.4 \pm 11)\times 10^{-9}$ \\
 $\bs^{XY}$ 	 & $(0.2 \pm 3.9)\times 10^{-9}$\\
 $\bsa^{A}$ 		 & $(-2.6 \pm 4.4)\times 10^{-9}$\\
 $\bsa^{B}$		& $(-0.3 \pm 4.5)\times 10^{-9}$ \\
 $\bs^{TX}$		& $(-3.1 \pm 5.1)\times 10^{-5}$ \\
$\bs^{TY}$ 		 & $(0.1 \pm 5.4)\times 10^{-5}$ \\
$\bs^{TZ}$ 		 & $(1.4 \pm 6.6)\times 10^{-5}$ \\
\hline
\end{tabular}
\end{table}

In our final analysis, we combine the three analysis with both the $\bar s^{\mu\nu}$ and $\ba{w}^{J}$ coefficients: (i) planetary ephemerides analysis given by Tab.~\ref{tab:results} with the correlation matrix from Tab.~\ref{tab:corr} (or equivalently the results from Tab.~\ref{tab:results} on the linear combinations given by Eqs.~(\ref{eq:lin_ind})), (ii) LLR data analysis from~\cite{battat:2007uq} summarized in Tab.~\ref{tab:LLR} with linear combinations given by Eqs.~(\ref{eq:bsla}-\ref{eq:bslb}) and~(\ref{eq:lin_LLR2}) and (iii) atom interferometry gravimetry analysis from \cite{muller:2008kx,chung:2009uq} presented in Tab.~\ref{tab:atom} with the linear combinations given by Eq.~(\ref{eq:lin_at}). The (marginalized) results of this fit are presented in Tab.~\ref{tab:fitall}.

\begin{table}[htb]
\caption{Estimated mean and 1$\sigma$ uncertainty of the SME coefficients obtained with a fit combining results from Sec.~\ref{sec:analysis}, LLR data analysis from \cite{battat:2007uq} and atom interferometry gravimetry  experiment~\cite{chung:2009uq,muller:2008kx}.}
\label{tab:fitall} % is used to refer this table in the text
\centering
\begin{tabular}{c r  l}
\hline
SME coefficients &  \multicolumn{1}{c}{Estimation} \\\hline\hline
$\bs^{XX}-\bs^{YY}$			 & $(9.6 \pm 5.6)\times 10^{-11}$ \\
$\bs^Q=\bs^{XX}+\bs^{YY}-2~\bs^{ZZ} $ 	 & $(1.6 \pm 0.78)\times 10^{-10}$\\
$\bs^{XY}$        	 & $(6.5 \pm 3.2)\times 10^{-11}$\\
$\bs^{XZ}$ 		 & $(2.0 \pm 1.0)\times 10^{-11}$\\
$\bs^{YZ}$ 		 & $(4.1 \pm 5.0)\times 10^{-12}$\\\hline
$\bs^{TX}$	 	& $(-7.4 \pm 8.7)\times 10^{-6\phantom{1}}$ \\
$\bs^{TY}$		 & $(-0.8 \pm 2.5)\times 10^{-5\phantom{1}}$ \\
$\bs^{TZ}$ 		 & $(0.8 \pm 5.8)\times 10^{-5\phantom{1}}$ \\\hline
$\alpha\ba{e}^{X}+\alpha\ba{p}^{X}$	 	& $(-7.6 \pm 9.0)\times 10^{-6\phantom{1}}$ & GeV/c$^2$\\
$\alpha\ba{e}^{Y}+\alpha\ba{p}^{Y}$	 	& $(-6.2 \pm 9.5)\times 10^{-5\phantom{1}}$& GeV/c$^2$ \\
$\alpha\ba{e}^{Z}+\alpha\ba{p}^{Z}$	 	& $(1.3 \pm 2.2)\times 10^{-4\phantom{1}}$ & GeV/c$^2$\\
$\alpha\ba{n}^{X}$	 	& $(-5.4 \pm 6.3)\times 10^{-6\phantom{1}}$ & GeV/c$^2$\\
$\alpha\ba{n}^{Y}$	 	& $(4.8 \pm 8.2)\times 10^{-4\phantom{1}}$& GeV/c$^2$ \\
$\alpha\ba{n}^{Z}$	 	& $(-1.1 \pm 1.9)\times 10^{-3\phantom{1}}$ & GeV/c$^2$\\\hline
\end{tabular}
\end{table}

The resulting estimations do not show any significant deviations from GR. The combinations of the three data analyses allow one to estimate each of the coefficients individually. The spatial part of $\bs^{JK}$ is completely determined by the combination of planetary ephemerides and LLR data. The atom interferometry gravimetry is not accurate enough to provide any significative improvement on the uncertainty of these coefficients. With an improvement of 2 orders of magnitude, the atom gravimetry data would become significative to estimate the $\bs^{JK}$ coefficients.   On the other hand, the three datasets are required in order to decorrelate the $\bs^{TJ}$ and the $\ba{w}^J$ coefficients. The uncertainties on $\bs^{TJ}$ are much larger than those shown in Tab.~\ref{tab:resINPOPLLR} where the coefficients $\ba{w}^J$ have been neglected. This reflects the fact that the individual coefficients are still highly correlated.

\section{Discussion}\label{sec:dis}
First of all, the accuracy of the constraints on the SME coefficients obtained in Tab.~\ref{tab:results} (planetary orbital dynamics alone) are of the same order of magnitude as the binary pulsars~\cite{shao:2014qd} constraints on the SME coefficients with an improvement of one order of magnitude on the coefficients $\bs^{YZ}$. Nevertheless, it is known that nonperturbative effects (similar to those computed in \cite{damour:1993vn}) may arise in binary pulsar systems. The nonperturbative effects depend highly on the fundamental theory (for example, see \cite{yagi:2014jk,*yagi:2014fk} for nonpertubative calculations in Einstein-Aether theory or in Ho\v{r}ava gravity). In general, the results from \cite{shao:2014qd,shao:2014rc} are effective constraints on the strong field version of the $\bs^{\mu\nu}$ that may include nonperturbative strong field effects and one should be careful when comparing strong field tests and weak field tests as the one performed in Sec.~\ref{sec:analysis}. The results shown in Tab.~\ref{tab:results2} improve the current Solar System constraints~\cite{kostelecky:2011ly} by 1 to 3 orders of magnitude.  Furthermore, the analysis combining planetary orbital dynamics and LLR from Tab.~\ref{tab:resINPOPLLR} improves by 2 to 3 orders of magnitude the previous results that combined LLR and atom interferometry. This shows the high impact provided by planetary ephemerides analysis.

As mentioned in Sec.~\ref{sec:analysis}, our results show that the estimated SME coefficients are highly correlated. The correlations are due to the similarity of the orbital planes of all the planets. Therefore, one way to improve the results by reducing the correlations is to use bodies with different orbital planes like e.g. asteroids. This can be achieved for example with Gaia observations similar to what is proposed in \cite{mouret:2011uq}.

The constraints obtained in Sec.~\ref{sec:analysis} are mainly due to the internal planets. For instance, Jupiter has absolutely no influence on the results shown in Tab.~\ref{tab:results}. This is a consequence of its not so well-known orbit. An improvement by a factor 10 on the knowledge of Jupiter's orbit is required for that planet to play a significant role in this analysis. Therefore, the improvement of Jupiter's trajectory expected from the analysis of Juno's radioscience and very long baseline interferometry data~\cite{anderson:2004uq} may improve the result of our analysis. In particular, it will reduce some of the correlations which will lead to an improvement of the estimations of the SME coefficients. In the same spirit, the influence of Saturn is weak but nevertheless highly important to decorrelate the coefficients. Furthermore, an improvement of Mercury's orbit by a factor 10 (which can be regarded as the improvement by Messenger's data that are not yet included in INPOP10a analysis~\cite{fienga:2011qf}) will lead to an improvement on the estimations of $\bs^{YZ}$ by a factor 2 and to a 10\% improvement on the coefficients $\bS^{TY}$ and $\bS^{TZ}$  (but to no improvement at all on the other coefficients). In summary, the best way to improve the current analysis is to improve the trajectory of the ``badly'' determined planetary orbits in order to improve the decorrelation instead of improving more the planets that are already very well determined.

As mentioned in Sec.~\ref{sec:orbito}, the influence of the $\bs^{TT}$ and the $\ba{w}^T$ coefficients on the orbital dynamics only appears at the next post-Newtonian order and these coefficients are therefore not constrained by our analysis. Nevertheless, these coefficients will play an important role in the light propagation~\cite{bailey:2009fk,tso:2011uq}. Therefore, planetary ephemerides may potentially constrain this coefficient by considering the effect of $\bs^{TT}$ on the light time of the radioscience Range observables used in the analysis. Other opportunities to constrain this coefficient are to consider a conjunction experiment like the one performed with the Cassini spacecraft~\cite{bertotti:2003uq} (or to analyze Cassini data within the SME formalism as proposed in \cite{hees:2014ys,*hees:2014vn,*hees:2015zr}) or to consider Very Long Baseline Interferometry observations similar to what has been done for the $\gamma$ post-Newtonian parameter~\cite{lambert:2009bh,*lambert:2011yu}. 

The multiplication of the numbers of SME coefficients that need to be considered leads to an increase in the uncertainties on each individual coefficients. This is due to the correlations between the different coefficients that appear when their number is increased. Therefore, it is highly important to increase the number of analyses to constrain SME. In this communication, we have shown how a combination of three analyses can disentangle the different coefficients. Nevertheless, the coefficients shown in Tab.~\ref{tab:fitall} are still highly correlated, especially in the $\ba{w}^J$ sector. One way to reduce these correlations is to use more observations that are sensitive to other combinations of the $\ba{w}^J$ coefficients. This can be done in two ways: (i) to consider different source bodies that generate the gravitational field and (ii) to use more orbital geometry like e.g. asteroids dynamics as already mentioned. The first point is related to the fact that the $\ba{w}^J$ coefficients enter the equations of motion  essentially through the properties of the source body. In this communication, only two source bodies have been used: the Sun (in the planetary orbital dynamics analysis) and the Earth (in LLR and in atom interferometry gravimetry). Considering more source bodies with different compositions can help to reduce correlations. In this sense, a test using the satellites around the different planets would be highly relevant.

Finally, we would like to soften the results presented here. First of all, we insist on the fact that the constraints obtained in Sec.~\ref{sec:analysis} correspond to the intervals in which the differences of INPOP10a postfit residuals are below 5 \%, as they are obtained directly from the limits of Tab.~\ref{tab:inpop} coming from~\cite{fienga:2011qf}. As such, they do not directly represent the usual 1$\sigma$ confidence interval. A cleaner approach would be to include the SME equations of motion directly in the planetary ephemerides software and to estimate the SME coefficients directly from the raw data, which corresponds to the approach usually used for estimating the PPN coefficients~\cite{verma:2014jk,konopliv:2011dq,pitjeva:2013fk} or more recently to constrain the MOND theory~\cite{hees:2014jk}. Our analysis demonstrates the impact of such an analysis and therefore, provides a strong incentive.

In addition, the LLR data analysis has been performed by fitting some oscillating signatures in the LLR data residuals. This approach is not optimal since it suffers from two drawbacks. First, the oscillating signatures derived in \cite{bailey:2006uq} have been computed analytically using several approximations. They can be used to estimate an order of magnitude on the different effects produced by SME but they are not optimal for a real data analysis (furthermore, the signatures used in \cite{bailey:2006uq} include only the dominant oscillations, several other frequencies are produced by SME and ignored in the data analysis). Second, fitting in the residuals is not optimal since it does not allow one to analyze the correlations between the SME coefficients and the other parameters that are usually fitted in a standard LLR data analysis. For these reasons, a cleaner analysis would include the SME equations of motion directly in the software used to reduce LLR data. Results obtained in \cite{battat:2007uq} and in this communication give strong motivations to perform such an analysis.

Finally, the atom interferometry gravimetry analysis should be interpreted with caution.  The atom interferometry gravimeter results from~\cite{muller:2008kx,chung:2009uq} assume a model of the local solid Earth tides.  While such models can be partly analytically based, it is known that the many frequencies of the Earth tides include all of the frequencies in the SME signal~\cite{tamura:1987hc}. If any aspect of the tidal model includes fitting sinusoidal functions to local gravimetry measurements or global measurements of, for example, the ocean heights~\cite{egbert:1994jk}, the signal for the SME may be partly subtracted due to the strong correlation with the tidal signal.

\section{Conclusion}
In this communication, we have shown that the planetary orbital dynamics allow one to constrain a violation of  Lorentz symmetry with an impressive accuracy. In Sec.~\ref{sec:analysis}, we use the current limits on supplementary advances of perihelia and nodes provided by INPOP10a~\cite{fienga:2011qf} to estimate the SME coefficients $\bs^{\mu\nu}$ and $\ba{w}^J$. In this analysis, the coefficients $\bar c^{\mu\nu}$ have been neglected since they are already constrained with a high level of accuracy~\cite{kostelecky:2011ly} but they can be considered in a future work. Our analysis has been performed using a standard Bayesian inversion. Results on the SME coefficients are given in Tab.~\ref{tab:results}. No significative deviation from GR is observed. As mentioned in Sec.~\ref{sec:analysis}, these estimations are highly correlated (see Tab.~\ref{tab:corr} or Fig.~\ref{fig:hist1}). We have identified numerically the linear combinations of the SME coefficients that can be estimated independently from planetary ephemerides. The estimations on these combinations are given in Tab.~\ref{tab:results2}. These two results are completely equivalent (as long as one uses the correlation matrix with the first estimation). Our results produce uncertainties similar to those obtained from binary pulsars data \cite{shao:2014qd,shao:2014rc} on most of the coefficients and improve the constraints on $\bs^{YZ}$ by one order of magnitude. Moreover, we improve the current best weak field tests by 2 to 3 orders of magnitude.

We also perform a combined estimation of the SME coefficients using results from three different analyses: (i) the planetary ephemerides analysis performed in Sec.~\ref{sec:analysis}, (ii) the LLR data analysis performed in \cite{battat:2007uq} and (iii) the atom interferometry gravimetry analysis realized in \cite{muller:2008kx,chung:2009uq}. The combination of LLR and planetary ephemerides leads to the best current estimations on the pure gravity SME coefficients as shown in Tab.~\ref{tab:resINPOPLLR} (when neglecting the $\ba{w}^J$ coefficients). In these three analyses, we also take into account potential effects produced by a Lorentz violation in the matter-gravity coupling which is parametrized by the $\ba{w}^J$ coefficients. Finally, the combination of the results from the three data analyses leads to the first independent estimations of the $\bs^{\mu\nu}$ and $\ba{w}^J$ coefficients. The results are presented in Tab.~\ref{tab:fitall}. The obtained uncertainties are relatively large, which is due to the numbers of coefficients considered and to the remaining correlations. Some ideas to reduce these correlations are proposed in Sec.~\ref{sec:dis}.

\begin{acknowledgments}
A.H. acknowledges support from ``Fonds Sp\'ecial de Recherche" through a FSR-UCL grant, thanks A. Fienga for interesting explanations about the estimations of the supplementary nodes and perihelia with INPOP planetary ephemerides and thanks N. Mohapi for interesting discussions on the Bayesian inversion. Q.~G.~B. is supported by the NSF Grant No. PHY-1402890. Q.~G.~B., C.~G. and P.~W. acknowledge financial support from Sorbonne Universit\'es through an ``Emergence'' grant. C.L.P.L. is grateful for the financial support of CNRS/GRAM and ``Axe Gphys'' of Paris Observatory Scientific Council.
\end{acknowledgments}

\bibliographystyle{apsrev4-1}
\bibliography{../../../JPL/JPL_byMe/biblio}
%\bibliography{SME}
\end{document}